# The European Quantum Technologies Roadmap

## Introduction

Max F. Riedel[1], Tommaso Calarco[1]

Within the last two decades, Quantum Technologies (QT) have made tremendous progress, moving from Noble Prize award-winning experiments on quantum physics [Nobel] into a cross-disciplinary field of applied research. Technologies are being developed now that explicitly address individual quantum states and make use of the "strange" quantum properties, such as superposition and entanglement. The field comprises four domains: Quantum Communication, where individual or entangled photons are used to transmit data in a provably secure way; Quantum Simulation, where well-controlled quantum systems are used to reproduce the behavior of other, less accessible quantum systems; Quantum Computation, which employs quantum effects to dramatically speed up certain calculations, such as number factoring; and Quantum Sensing & Metrology, where the high sensitivity of coherent quantum systems to external perturbations is exploited to enhance the performance of measurements of physical quantities.

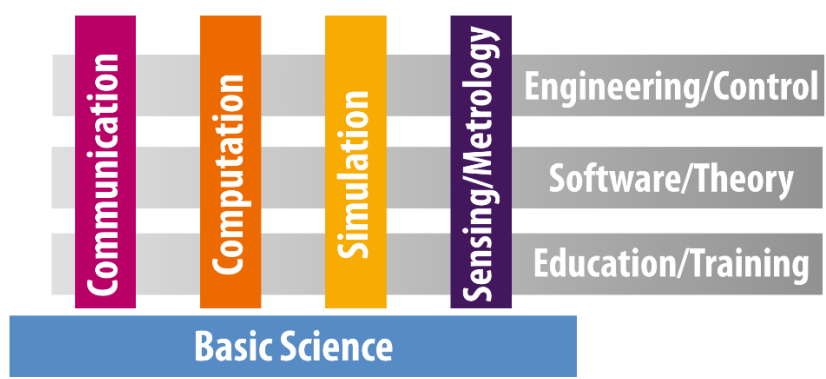

*Figure 1: Structure of the research field of Quantum Technologies, according to final report of the High-Level Steering Committee on the European QT Flagship initiative [HLSC]*

Recently QT have received a lot of public attention: Governments have launched large research programmes on QT, such as the Chinese programme (which includes the launch of a satellite and the instalment of a quantum key distribution link between Beijing and Shanghai) or the European Quantum Technologies Flagship initiative, summing up to several billion Euros of public funding for the field worldwide. At the same time, large multinational companies, including Google, IBM, Intel, Microsoft and Toshiba, have started to invest heavily in QT, especially in Quantum Computing and Quantum Communication. Also, a number of start-up companies were established during the last decade which successfully offer QT to specialized markets.

One success factor for the rapid advancement of QT is a well-aligned global research community with a common understanding of the challenges and goals. In Europe, this community has profited from several EC funded coordination projects, which, among other things, have coordinated the creation of a 150-page

[1] University Ulm and Center for Integrated Quantum Science and Technology (IQST), Albert-Einstein-Allee 11, D-89081 Ulm, Germany

QT Roadmap [Roadmap16]. This article presents an updated summary of this roadmap. Besides sections on the four domains of QT, we have included sections on Quantum Theory & Software and on Quantum Control, as both are important areas of research that cut across all four domains. Each section, after a short introduction to the domain, gives an overview on its current status and main challenges and then describes the advances in science and technology foreseen for the next ten years and beyond.

## References and footnotes

# Quantum Communication

Rob Thew[2], Nicolas Gisin[2]

## Introduction

Quantum communication involves the generation and use of quantum states and resources for communication protocols. Its main applications are in provably secure communication, long-term secure storage, cloud computing and other cryptography-related tasks, as well as in the future, a secure "quantum web" distributing quantum resources like entanglement, nonlocality, randomness and connecting remote devices and systems. Typically, the underlying protocols are built on quantum random number generators (QRNG) for secret keys and quantum key distribution (QKD) for their secure distribution. The archetypal QRNG involves a photon impinging on a beam splitter followed by two detectors associated to the bit values 0 and 1, where the origin of the randomness is clearly identified. QKD systems take this one step further to distribute this randomness in a correlated way; such that two parties share the same random string in a private and secure fashion. Secure solutions based on quantum encryption are importantly also immune to attacks by quantum computers, and are commercially available today, as is quantum random number generation. Indeed, recently it has been shown that the camera in mobile phones can be used as a QRNG, opening the door to potentially massive commercial opportunities.

## Current Status

Currently, typical fibre-based QKD systems can only function over distances of around 100km for commercial systems, although academic prototypes can push this to around 300km [Korzh15], which is limited by transmission loss in optical fibres; quantum information is secure because it cannot be cloned, but for the same reason it cannot be relayed through conventional repeaters. In classical optical telecommunication, the problem of loss is solved by using simple optical amplifiers that restore the transmitted signal. However, these are of no use for quantum communication as they are intrinsically noisy and create so many errors that any quantum key being transmitted would not survive. So, quantum communication must reinvent the repeater concept, using quantum hardware that preserves the quantum nature, the entanglement. Therefore, repeaters based on trusted nodes or fully quantum devices, possibly involving satellites, are needed to reach global distances. Trusted node relays consist of multiple QKD systems that are chained together to build longer and more complex fibre networks, which can provide backbone or access [Frohlich13] network architectures but require a trusted environment where the devices can be connected together. Satellites [Scheidl13] and high altitude platform stations (HAPS), which include drone-based scenarios, provide an alternative approach and potentially complimentary solution. Fully quantum-secure solutions for long-distance quantum networks, based on quantum repeaters exploiting multimode quantum memories, aim to increase the distances between trusted nodes as well as providing the ability to distribute entanglement to distant locations for interfacing with quantum processors or sensors and provide opportunities for novel applications. Quantum repeaters [Sangouard09] allow one to break the transmission distance up into shorter distances where entanglement can be prepared and stored in a quantum memory – a device capable of storing quantum states. Once the different sections are ready they can be connected by so-called Bell-state measurements until the entire communication length is entangled, for example, allowing one to 'teleport' qubits directly to their destination, thus avoiding transmission losses. There is currently enormous activity in developing quantum memories using a wide variety of physical platforms [Bussieres13] that are both efficient (information is not lost) and offer scalable solutions for the grand challenge of continental and global scale quantum-secure communication and entanglement distribution.

---

[2] Group of Applied Physics, University of Geneva, CH-1211 Geneva 4, Switzerland

There is also currently a great deal of theoretical work taking place on developing new protocols and new approaches to certifying systems, for example, their security. This work on new protocols and certification takes several different approaches from work bringing quantum and classical security experts together [Buchmann17] to developing practical security proofs, or those coming from a more fundamental perspective, i.e. studies of nonlocality in what are called "device-independent" protocols [Acin16], or related "self-testing" strategies. Certification is also starting to take into account commercial considerations to have devices and systems certified for compliance with industry standards. Standards themselves represent an important challenge that has begun to be addressed by working groups that bring together industry and academics, as well as national metrology labs [MIQC2].

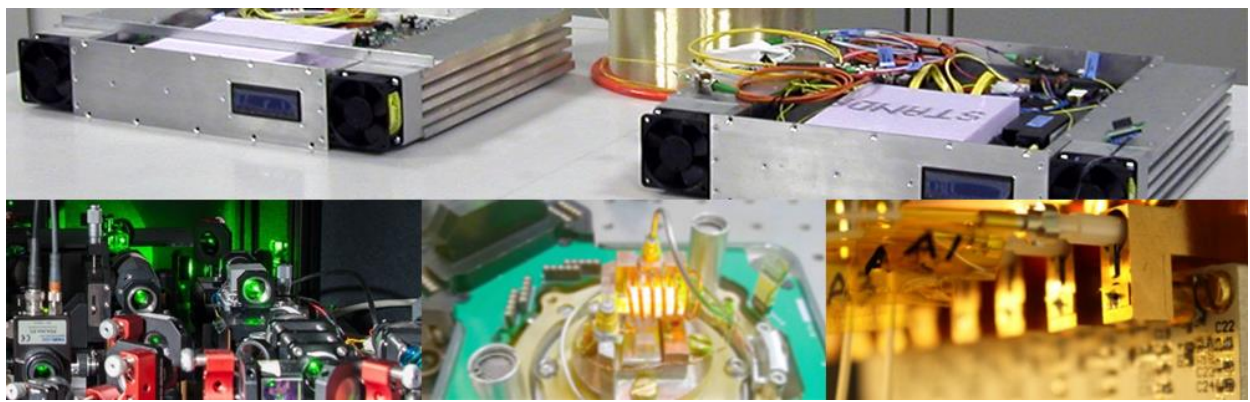

*Figure 2: Photo montage of (a) University prototype QKD system, (b) Entangled photon pair source, (c) Solid state (rare-earth ion) quantum memory for quantum repeaters, (d) Array of superconducting nanowire single photon detectors.*

## Advances in science and technology needed to meet future challenges

Quantum communication is both a broad field, addressing numerous tasks and applications, but also one that spans research and engineering challenges from fundamental to applied and towards the development of prototype devices and systems as well as managing their functionality in diverse network architectures. It is also a field in which there is an incredible range of possible technology platforms that can be exploited for any given task. As such, we will not give a detailed roadmap of what is required in all of these different platforms, but focus on the current and future challenges that are being targeted.

Foreseeable within the next three years is the development of autonomous QKD systems over metropolitan distances that will address low deployment costs, high secure key rates (> 10Mbps) and multiplexing. It is expected that integrated photonic solutions will be critical in these efforts. Certification and standards for quantum communication devices and systems will be established, as required by the security community, industry, ESA and government organisations. QRNGs e.g. for use as components of cheap devices will be developed targeting high-volume markets or high-speed systems, including entropy source and application interface. QRNG and QKD devices and systems will address issues of practicality, compactness, high-rates, or include novel approaches that address security vulnerabilities or certification challenges. To extend QKD beyond the direct communication distance limit (> 500km), the underlying technologies for trusted nodes, quantum repeaters, HAPS and satellites will need to be developed. Quantum repeater and multipartite entanglement-based network building blocks are aiming to demonstrate improved performance for core technologies, including: efficient and scalable quantum memories and interfaces; frequency conversion; teleportation; entanglement distillation; error correction; sources of single photons and entanglement, and detectors. Practical protocols and efficient algorithms for quantum networks, e.g. digital signatures, position based verification, secret sharing, oblivious data searching, will be developed. Solutions that use both classical and quantum primitives will also be explored to ensure compatibility with existing infrastructure as well as working towards long-term, future-proof, security.

In 6 years, we will likely see QKD in test-bed networks, demonstrating long distances via trusted-nodes, HAPS or satellites, as well as multi-node or switchable intra-city networks, all of which will require large-scale infrastructure projects to be initiated. Autonomous QKD systems suitable for low-cost volume manufacturing as well as systems targeting increased ($> 100 \text{Mbps}$) secure key rates over metropolitan distances will be targeted. Quantum repeaters and entanglement-based networks beating direct communication distances will be demonstrated. Hardware and software for entanglement-based networks will be developed, including multipartite and device-independent-inspired protocols, with explicit and demonstrable assumptions about security, e.g. for QRNG as well as QKD over $> 10 \text{km}$.

In the long-term it is important to consider not only the research but the innovation environment that will have been created and what will follow. The long-term objectives of the quantum communication community include: generalised use of autonomous QKD systems and networks; device independent QRNG systems and QKD over metro-area distances; quantum cryptography over $> 1000 \text{km}$, and protocol demonstrations, e.g. cloud computing, on photonic networks connecting remote quantum devices or systems.

To ensure the success of all of these objectives there is a need for dedicated engineering support for all of these activities across the research and development spectrum. These engineering, as well as control, solutions are aiming to enable scaling and volume manufacturing, e.g. development of high-speed electronics and opto-electronics, including FPGA/ASIC, integrated photonics, packaging, compact cryo-systems, and other key enabling technologies, to provide solutions compatible with operating in existing communication networks. This will also need support in terms of theory and software development of protocols and applications that build on, or go beyond, standard QRNG- and QKD-based primitives, as well as novel approaches for their certification, including methods to test and assess the performance of quantum networks; more efficient algorithms and security proofs targeting practical systems, including the combination of classical and quantum encryption techniques for holistic security solutions and expanding the potential application market.

## Conclusion

The security of our information-based society is of rapidly increasing importance. The long-term secure management of data in transit and at rest is of paramount importance for society and the economy as well as our infrastructures and services, our prosperity, as well as for political stability. These risks are augmented by growing technological threats such as the development of a quantum computer, which makes the most commonly used asymmetric cryptography algorithms vulnerable and poses a systemic threat to long-term security. Quantum communication provides solutions that can be integrated into existing infrastructure and protocols as well as opening up new application regimes. These ambitious objectives, and the innovative environment being developed to realise them, should form a solid basis to ensure that Quantum Technologies play a leading role in the science, technology and digital economy of the 21ˢᵗ century.

We thank the many members of the community who have contributed to the content of this article, in particular P. Grangier, R. Renner, G. Ribordy, A. J. Shields, and R. Ursin.

# Quantum Computation

Frank K. Wilhelm[3], Daniel Esteve[4], Christopher Eichler[5], Andreas Wallraff[5]

## Introduction

A quantum computer based on the unitary evolution of a modest number of robust logical qubits (N>100) operating on a computational state space with $2^N$ basis states would outperform conventional computers for a number of well identified tasks. A viable implementation of a quantum computer has to meet a set of requirements known as the DiVincenzo criteria: That is, a quantum computer operates on an easily extendable set of well characterized qubits (1) whose coherence times are long enough for allowing coherent operation (2) and whose initial state can be set (3). The qubits of the system can be operated on logically with a universal set of gates (4) and the final state can be measured (5). To allow for communication, stationary qubits can be converted into mobile ones (6) and transmitted faithfully (7). It is also understood that it is essential for the operation of any quantum computer to correct for errors that are inevitable and much more likely than in classical computers.

Today quantum processors are implemented using a range of physical systems. Quantum processors operating on registers of such qubits have so far been able to demonstrate many elementary instances of quantum algorithms and protocols. The development into a fully featured large quantum computer faces a scalability challenge which comprises of integrating a large number of qubits and correcting quantum errors. Different fault-tolerant architectures are proposed to address these challenges. The steadily growing efforts of academic labs, startups and large companies are a clear sign that large scale quantum computation is considered a challenging but potentially rewarding goal.

**Toward scalable architectures for the gate model.** Controlling and error-correcting the unitary evolution of about 100 logical qubits will be a major milestone in the quest for overcoming present-day classical processors on specific tasks, e.g. in quantum chemistry or simulation. Realizing logical qubits includes encoding in a larger number of physical qubits with sufficient functionality in a viable architecture. This may imply, for example, creating large scale 2D traps for ions or realizing the surface code architecture for superconducting qubits. The most promising architectures for achieving fault tolerance may be specific to the respective platform but address common challenges.

**Alternative architectures for quantum computing.** Given the significant challenges of implementing fault-tolerant gate-based processors, alternative concepts subject to different sets of challenges are actively pursued. Most prominently, quantum assisted annealing is followed by companies such as D-Wave Systems, Google and a number of academic labs, with quantum speedup being unclear at best.

In the following, we will describe the current status and the advances in science and technology needed to meet the challenges for the five most important QC platforms.

---

[3] Saarland University, D-66123 Saarbrücken, Germany
[4] Service de Physique de l'Etat Condensé, CEA-Saclay, 91191 GIF-SUR-YVETTE, France
[5] ETH Zurich, Department of Physics, CH-8093 Zürich, Switzerland

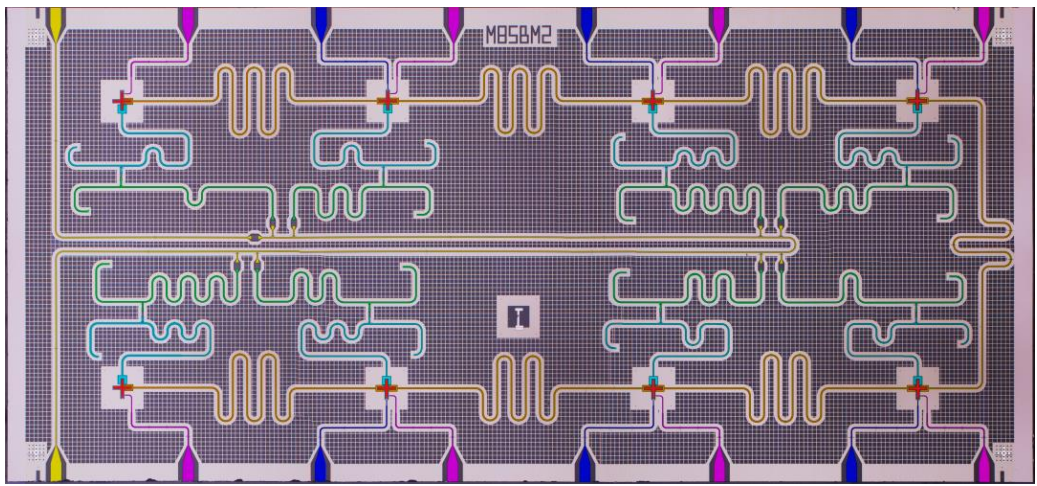

*Figure 3: False-coloured image of an 8-qubit superconducting quantum processor fabricated at ETH Zurich. All eight qubits (red) are measured using a common readout line (yellow), by coupling each qubit (red) to a pair of readout resonator (cyan) and Purcell filter (green). Qubit control is enabled by individual charge lines (purple) and flux lines (blue). Coupling between nearest neighbour qubits is mediated by bus resonators (orange).*

## Status and challenges

**Trapped ions** [Blatt08, Monroe13, Home09]. Ion trap quantum computing typically operates on a qubit register formed by a linear string of ions confined in a Paul trap. Each physical qubit is based on two internal levels of a single ion - defined within a Zeeman or hyperfine manifold or corresponding to a forbidden optical transition. Gate operations use microwave or laser fields.

Quantum algorithms have been performed on strings of up to seven ions confined in a linear trap. Longer chains of up to 20 ions and 2D crystals of up to ~300 ions have been trapped and used for quantum state engineering or simulation. Individual qubits can be initialized with error below ~$10^{-3}$, are controlled with gate errors of ~$10^{-6}$, and read out with an error of ~$10^{-4}$. Two-qubit gates have errors of ~$10^{-3}$. The conversion from stationary to flying qubits has been demonstrated, as well as the transfer of quantum information over short distances by physically transporting ions across a microchip.

Scalability remains the most significant challenge in ion systems for which well-defined approaches based on micro-fabricated traps and photonic interconnects are developed. Various fabrication techniques and electrode configurations are investigated. Micro-fabricated 2D RF-trap arrays have already been successfully demonstrated. A difficulty encountered in miniaturized ion traps is the marked growth of the electric-field noise in the vicinity of trap surfaces causing unwanted motional heating. This issue has been addressed by operating at cryogenic temperatures, and/or by applying an in-situ cleaning of the trap surface.

Further short- and mid-term goals specific to ions in microfabricated traps include demonstration of high-fidelity gates in multiple ion registers, operation of 2D traps, integration of optics and control electronics and demonstration of high-fidelity quantum information transport between ion registers, and between three or more networked traps.

**Superconducting circuits** [Nakamura99, Devoret13]. Quantum computation with superconducting circuits exploits the intrinsic coherence of superconductors and the Josephson effect as a resource of dissipationless non-linearity for making artificial atoms. Qubits are realized as resonant microwave circuits embedding a Josephson tunnel junction, of which the two lowest energy levels are used as an effective quantum bit. Superconducting qubits are fabricated with thin-film technology, are probed and controlled with microwave radiation and can be strongly coupled to each other by circuit elements. Superconducting resonators and cavities provide opportunities for coupling widely different types of qubits in hybrid devices, including atoms, ions and impurity spins in quantum dots, crystals, and microtraps.

Industry interest in superconducting quantum computing has sharply risen in recent years illustrating its potential. Processors with 4-17 qubits have demonstrated the basis of quantum error correction protocols, elementary quantum algorithms, and simulations. Universal gate operations are performed with fidelities in excess of 99.9% for single qubits and 99.5% for two-qubit gates. The use of optimized parametric amplification routinely enables single-shot, non-demolition qubit measurements with fidelities exceeding 99%. The coherence times of qubits are constantly increasing. At the same time, fast classical control electronics, as required for real-time feedback, are rapidly advancing.

Designing and fabricating large scale superconducting circuits addressing all circuit elements without crosstalk is challenging. Microfabricated superconducting qubits are sensitive to imperfections in their fabrication limiting yield and reproducibility of device parameters. Both aspects require optimization of design and production processes. Operation of devices below 50 mK requires refrigeration technology which is realizable beyond a few hundred. Goals include to realize an extensible quantum processor architecture, allowing copy-pasting of unit cells, develop transitioning from millimetre to centimetre scale chips, and from lateral to vertical coupling of all control signals to the chip, realize an extensible, control electronic architecture for control of the quantum circuit, operating either at room temperature, cryogenically, or a combination of both, and develop automated tune-up and calibration procedures.

**Electronic semiconductor qubits.** In semiconductor host materials single electrons can be either trapped by isolated donor atoms, confined in ultra-small islands or using gate-defined potentials, or by topological effects. The spin degree of freedom in these systems is considered promising due to its long coherence time. These devices can be measured and controlled fully electrically and their fabrication exploits the same technologies as the semiconductor industry. Recently, group IV materials such as Si/SiGe have attracted increasing attention, as they offer long spin coherence times when using nuclear spin-free $^{28}$Si isotopes.

Quantum dot circuits [Loss98] with up to five quantum dots have been controllably loaded. Single qubit gates have fidelities in excess of 99%, spin states are initialized with 99.9% fidelity, and single shot readout of up to three qubits was demonstrated with an average fidelity of 97%. Coherence times as long as T2 (T2*) = 500 (0.2) ms have been measured in isotopically enriched $^{28}$Si. Coherent exchange coupling and interaction between two spins in a double dot have been demonstrated.

One of the main challenges remains the development and improvement of high fidelity two-qubit gates, particularly for donor spins. Various material needs to be investigated and eliminated. Further goals contain the 'Unit cell' demonstration of a scalable 2D spin qubit architecture, identification of robust and secure sources for high-purity semiconductor materials and demonstration of precise positioning of donor arrays.

**Impurity spins.** Atomic and molecular spins in solids such as colour centres, rare earth ions, deep donors, and molecular magnets, can use both the electron and nuclear spin degrees of freedom as qubits [Kane98]. Control of these systems is typically achieved by combining highly advanced techniques from NMR with optical manipulation. These systems promise good shielding from the environment leading to long coherence time.

The most advanced platform so far are nitrogen vacancy centres in diamond [Hanson08]. Initialization and single shot spin readout are achieved with optical control, while single qubit gates employ microwave fields. Two-qubit gates between multiple spins are based either on magnetic dipolar interactions or on long distance optical coupling. Multipartite entanglement, quantum teleportation over long distances, quantum error correction, and elementary quantum algorithms have been demonstrated. Despite recent progress, nano-positioning and the creation yield of defects is still a major and most pressing challenge.

**Linear Optics.** Linear-optical quantum computing (LOQC) employs single photons, linear optics elements (discrete or on chip), photon-counting measurements, and feed-forward but avoids using direct photon interactions in nonlinear media. To date, there are two main physical architectures for LOQC: The scheme by Knill, Laflamme and Milburn (KLM) [Knill01], and the one-way quantum computing scheme. The KLM scheme is based on the preparation of multi-particle entangled states and (entangling) multi-particle projective measurements. One-way quantum computing [Raussendorf01] exploits a series of adaptive single-qubit rotations and measurements applied to cluster states that provide the resource.

The control of large entangled states has been achieved experimentally. Small-scale algorithms have been demonstrated, including alternative computational models based on quantum walks. Complete architectures for LOQC still need to be developed and hard bounds on the required performance of photonic components have to be investigated theoretically.

## Conclusion

Many implementations of quantum information processors share common goals. Improving coherence properties of qubits and enabling to enhance single and two-qubit gate fidelities, at least beyond the fault tolerant threshold, is a goal pursued throughout. Within the next five years, demonstrations of error-corrected logical qubits with performance beyond the constituent physical qubits is expected in a few implementations, as well as fault-tolerant gates. To operate systems of many physical qubits in an extensible fashion, scalable classical control electronics and tune-up routines for large-scale quantum systems are to be realized. In five to ten years, demonstrations of quantum algorithms operating on logical qubits in a universal quantum computer are envisaged. At the same time functional quantum interfaces for short, medium and long distance communication between quantum computing modules are foreseen to be functional. On the time scale of ten years and beyond the demonstration of large scale quantum computation systems is pursued. With such systems solving technologically relevant problems, is expected to be feasible.

We thank the many members of the community who have contributed to the content of this article.

# Quantum Simulation

Jens Eisert[6], Immanuel Bloch[7], Maciej Lewenstein[8], Stefan Kuhr[9]

## Introduction

The idea of quantum simulation goes back to Richard Feynman, who suggested that interacting quantum systems could be efficiently simulated employing other precisely controllable quantum systems, even in many instances in which such a simulation task is expected to be inefficient for standard classical computers [Feynman82]. In general, the classical simulation of quantum systems requires exponentially large resources, as the dimension of the underlying Hilbert space scales exponentially with the system size. This scaling may be significantly altered by employing appropriate representations of the quantum state valid in specific situations. Similarly, solutions of certain classical optimization problems, in particular NP-hard and NP-complete ones, require exponential resources. Numerical methods, such as tensor networks or the density-matrix renormalization group (DMRG) approach, as well as Quantum Monte Carlo sampling allow for computing of ground state properties in certain situations. Such classical simulation methods are generally applicable to restricted classes of problems and have their limitations. For example, the systems sizes that can be studied numerically on classical computers are often rather small and it seems unlikely that these classical tools will be powerful enough to provide a sufficient understanding of the full complexity of many-body quantum phenomena. In the language of complexity theory, approximating the ground-state energy of local Hamiltonian problems is QMA-hard, and time evolution under local Hamiltonians is BQP-complete, so both amount to computationally hard problems. Similarly, finding a ground-state energy of a classical spin glass, or solving the travelling salesman's problem, are computationally difficult. Quantum simulators promise to overcome some of these limitations.

## Current Status

In 1982, Richard Feynman not only introduced the basic idea of a quantum simulator in his published script of a keynote speech, but discussed sophisticated notions of simulation times and notions of simulation, and even delineated blueprints for potential architectures [Feynman82]. This basic idea was further substantiated by work showing that a universal quantum computer would indeed be able to efficiently keep track of the dynamics of any local quantum system, allowing for precise error analysis by means of the Trotter formula [Lloyd96]. Since then, the research field of quantum simulation has been flourishing and developing into a core field within quantum information processing in its own right, addressing notions of simulating complex quantum systems in several readings and ramifications. A working definition of a quantum simulator can be given as follows: A quantum simulator is any physical quantum system precisely prepared or manipulated in a way aimed at learning interesting property of an interacting complex quantum or classical system. More specifically:

- A quantum simulator is an experimental system that mimics an interacting quantum system with many degrees of freedom (from condensed-matter, high-energy physics, cosmology or quantum

---

[6] Dahlem Center for Complex Quantum Systems, Freie Universität Berlin, D-14195 Berlin, Germany
[7] Fakultät für Physik, Ludwig-Maximilians-Universität München, Schellingstrasse 4, D-80799 Munich, Germany and Max-Planck-Institut für Quantenoptik, Hans-Kopfermann-Str. 1, D-85748 Garching, Germany
[8] ICFO-Institut de Ciencies Fotoniques, The Barcelona Institute of Science and Technology, 08860 Castelldefels (Barcelona), and ICREA, Passeig de Ll. Campanys, 23, 08010 Barcelona, Spain
[9] University of Strathclyde, Department of Physics, SUPA, Glasgow G4 0NG, United Kingdom

chemistry). Alternatively, it may serve to encode hard classical constrained optimization problems (such as satisfiability).

- The simulated models should address a challenging problem and further our understanding in the addressed field.
- The simulated models should be expected to be computationally intractable or difficult for classical computers.
- A quantum simulator should allow for broad control of the parameters of the simulated model, as well as for control of the preparation, manipulation and detection of the states of the system. This feature can then be used to test models and hypothesis over a wide parameter regime in a precise fashion.

It can be helpful to be able to set the parameters of the quantum simulation in such a way that the model becomes tractable using classical simulations for purposes of validation through known 'reference results'. At the same time, it should be clear that the certification of a quantum simulator does not necessarily require the efficient classical simulation of certain parameter regimes.

Before turning to architectures for quantum simulation, it is helpful to be reminded of classical simulation methods aimed at computing properties of quantum many-body systems. The new research field "Hamiltonian complexity" aims to identify obstacles that any such classical simulation must ultimately face: For example, approximating the ground-state energy of an interacting local Hamiltonian problem to polynomial accuracy in the number of particles is QMA-hard, limiting the hopes that a universal classical simulation of key models in condensed-matter physics could be achieved. Similarly, many classical complex optimization problems are proven to be NP-hard. Still, for many practical purposes, classical simulations of quantum and classical systems, including solving constrained optimization problems are possible for specific models and in many regimes, at least to the level of a heuristic understanding.

The term quantum simulator refers to a number of closely related concepts of devices that aim at simulating complex quantum systems, using other highly controlled quantum systems. One distinguishes

- **static quantum simulators** [Bloch12, Lewenstein12, Georgescu14], probing static properties of interacting systems such as ground-state features, from
- **quantum annealers** [Albash15] approximating solutions to classical optimization problems, employing quantum annealing/adiabatic methods, and
- **dynamical quantum simulators** [Lloyd96, Bloch12, Trotzky12], probing properties related to non-equilibrium [Eisert15].

In terms of how the simulation is performed, one discriminates

- **digital quantum simulators** [Feynman82, Lloyd96, Blatt12], which are based on quantum circuits implemented on a quantum computer, and may in principle be made fault tolerant,
- **analogue quantum simulators,** simulators that reconstruct the time evolution of an interacting quantum system under precisely controlled conditions [Bloch12, Trotzky12, Lewenstein12].

The advantage of analogue quantum simulators is that a large number of constituents can be addressed and experimented with, even using architectures that are available with present technology. Quantum simulations thereby offer new insights into phenomena of complex quantum systems, with applications ranging from condensed matter physics over statistical physics, high-energy physics, cosmology and possibly even notions of energy transfer in biological systems. It is conceivable that quantum simulators can also help to interpret measurement results originating from sophisticated measurement techniques applied to real materials, e.g., 2D electronic spectroscopy or transport measurements. Due to the precise control over the Hamiltonian parameters, quantum simulators provide a deeper understanding of the effects of inter-particle interactions and their influence on the overall properties of the system and could therefore even be used in the quest to engineer materials with specialized properties. A first step in this endeavour is usually to identify the underlying model Hamiltonian, which is then probed by the actual quantum simulation.

There are a number of physical platforms that allow for controlled quantum simulations. Promising advances have been achieved in these different systems at different levels of maturity at the present stage. Experimental platforms [Georgescu14] for quantum simulation comprise

- ultra-cold atomic and molecular quantum gases, specifically systems of cold atoms in optical lattices or continuous systems confined by atom chips,
- ultra-cold trapped ions,
- polariton condensates in semiconductor nanostructures,
- circuit-based cavity quantum electrodynamics,
- arrays of quantum dots,
- Josephson junctions and superconducting qubits that already have commercial applications in quantum annealers, and
- photonic platforms

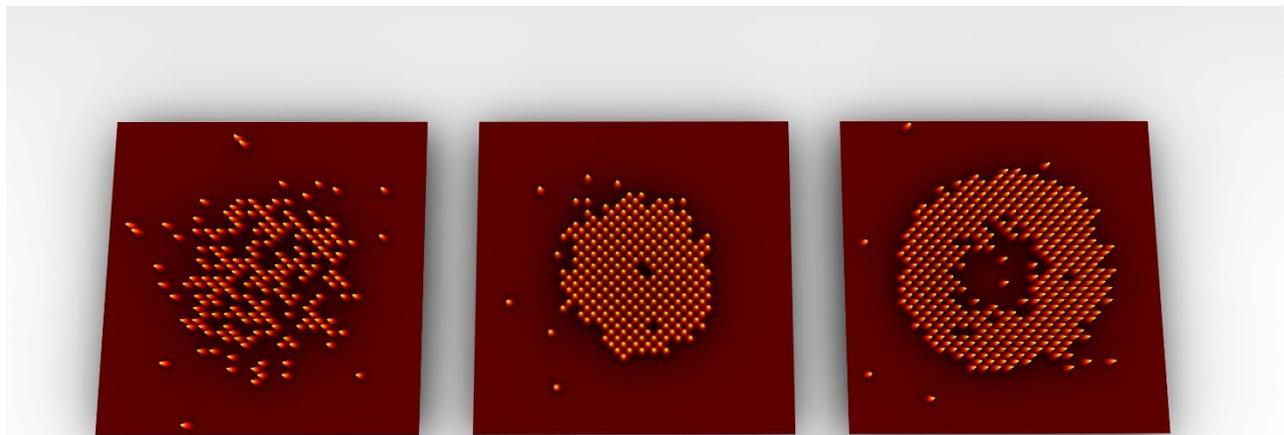

*Figure 4: Reconstructed quantum gas microscope images of single atoms held in an optical lattice. The images indicate two different phases of matter: a weakly interacting BEC (left) and a strongly interacting bosonic Mott insulator (middle/right) for two different atom numbers. Such single photographic snapshots of quantum matter enable to probe and analyse interacting many-body systems in completely new ways. (Source: Max-Planck Institute of Quantum Optics)*

## Advances in science and technology needed to meet future challenges

Quantum simulations allow to probe and explore properties of complex quantum systems under precisely controlled conditions. Despite significant advances both in theory and experiment, from the conceptual perspective, several problems remain open. This includes in particular the

- identification of models that are computationally difficult for classical simulations, and yet interesting and important from a physical point of view, the
- development of validation and verification tools for quantum simulators and classical simulation methods that can be used to capture the functioning of the quantum simulator in certain regimes and the
- design of experimental setups and implementations of sufficient size while at the same time exhibiting a high degree of control.

A key challenge is to find out whether the device has actually correctly performed the quantum simulation. This constitutes an important and intriguing problem in situations that are not classically attainable: The quantum simulator is performing tasks that one cannot efficiently keep track of, and still one would like to have evidence that the quantum simulator has functioned accurately. A commonly applied approach is to assume, that even if the entire family of models to be quantum simulated is inaccessible by classical means, there are suitable parameter regimes for which these models become fully or at least partially accessible for classical simulation. In some instances, the statements on the correctness of a quantum simulation can be made even without having to efficiently predict the outcome of the simulation.

However, there are some tasks in quantum simulation, such as approximating ground state energies, which not even a presumed quantum computer can overcome. Other aspects, such as the difficulty of computing long-time dynamics of many-body quantum systems, leave room for a computational quantum advantage of quantum simulators over classical ones, often eluded to as "quantum supremacy". Quantum annealers are believed to provide good approximations to solutions to NP-hard problems, but it is still unclear in what precise sense quantum simulators will provide an advantage over classical simulations [Albash15]. At the same time, another profound conceptual question arises: If error correction and fault tolerance are not available, it is still not fully understood to what extent verified quantum simulators and annealers can outperform classical computers.

## Conclusion

If a concise answer to this and related questions can be established, quantum simulators will play a pivotal role in our study of quantum many-body physics and allow to tackle the many complex challenges related to it. Moreover, even before these questions of verification and certification are completely resolved, which can reasonably be expected to be true within the next five to ten tears, analogue quantum simulators give us a novel tool to explore and understand features in interacting many-particle quantum systems and optimization problems that are beyond the reach of classical computers. As a long-term goal beyond the next ten years, it is expected that large-scale quantum simulations can be performed to tackle key questions in physics, materials science and quantum chemistry.

We thank the many members of the community who have contributed to the content of this article.

# Quantum Metrology, Sensing, and Imaging

Fedor Jelezko[10], Piet O. Schmidt[11], Ian Walmsley[12]

## Introduction

Measurement is the basis not only of science, which demands empirical quantitative assessment of phenomena, but also of commerce, which requires standards for metrology, without which there can be no common basis for the exchange of goods and services, including information. For these reasons, sensors are a vitally important technology, underpinning, for instance, navigation, geo-prospecting, chemical and materials analysis and characterization, fundamental science from the sub-nano to the galactic scale as well as determining the fundamental constants relied upon for industry and commerce.

The central concept of a sensor is that a probe interacts with an appropriate system, the properties of which are of interest, which changes of state of the probe. Measurements of the probe reveal the parameters that characterize the system. In quantum-enhanced sensors, the probe is generally prepared in a particular non-classical state. The encounter with the system typically modifies this state both usefully (by responding to the parameter of interest) and detrimentally (by erasing or decohering the probe). Properly designed measurements then determine in what way and to what degree the state of the probe has been altered by the encounter. This enables an estimate of the system parameters to be made, and thus the sensor response to be determined. The precision of this estimate as a function of the resources used (e.g. the number of particles in the probe or measurement time) is a measure of the effectiveness of the sensor. The best classical sensors exhibit a precision that scales proportionally to the square root of the number of particles N in the probe (known as the standard quantum limit, SQL) whereas the best quantum sensors can in principle attain a precision that scales as N (known as the Heisenberg limit).

Quantum enhanced sensing promises significant improvements in the precision with which properties of a wide range of systems can be estimated. The platforms for implementing new sensor protocols range from the nanoscale, by means of localized spins to the planetary scale, based on photons. Some platforms are already close to commercial application, others require new science and engineering to be fully viable. In the next sections we describe the current status and the advances in science and technology needed to meet the challenges for the most important quantum sensor platforms.

## Current status

### Atom and optical sensors

**Photonic sensors.** Practical designs for ultra-bright sources of quantum light with reduced noise [Valbruch16] and entanglement together development of novel principles for engineering practically useful quantum states and measurements [Kacprowicz10] have revolutionized photonic quantum sensing. For instance, recent demonstrations have shown the possibilities for multi-photon interferometry beyond the classical limit [Sluss17], and it has been shown that weak field homodyning could yield enhanced resolution in phase detection. Early experimental implementations of quantum ellipsometry indicated the high potential of quantum polarisation measurement while the first demonstration of quantum microscopy with NOON states demonstrated the potential of using fragile quantum states in imaging [Ono13]. In addition to quantum correlated photon states, (macroscopic) squeezed states of light can be also used as a resource for quantum-enhanced sensing. Currently squeezed light techniques are in use in GEO600, and will be adopted by LIGO [Chua14]. Squeezed light strategies are in development for deployment in a next-generation gravitational-wave detector, the Einstein Telescope. Squeezed light has also been exploited to resolve a small beam displacement, which in turn has been used to perform quantum-enhanced micro-rheology on a living cell [Taylor13].

---

[10] University Ulm and Center for Integrated Quantum Science and Technology (IQST), Albert-Einstein-Allee 11, D- 89081 Ulm, Germany

[11] QUEST Institute for Experimental Quantum Metrology, Physikalisch-Technische Bundesanstalt, D-38116 Braunschweig, and Institut für Quantenoptik, Leibniz Universität Hannover, D-30167 Hannover, Germany

[12] Clarendon Laboratory, Department of Physics, University of Oxford, Oxford OX1 3PU, United Kingdom

**Atomic sensors.** 2016 celebrates the 25th anniversary of atom interferometry, which harnesses the sensitivity of quantum superposition to create ultra-precise sensors for gravity, rotation, magnetic fields and time, surpassing their best classical counterparts. Owing to their maturity, they are ready for translation into commercial products. Sensors using micro-Bose-Einstein condensates enable exotic quantum states that allow precision sensing of fields near surfaces, for instance. Current atomic gravity sensors offer absolute measurements at the nano-g level or gravity gradient sensitivities surpassing a 100 pico-g change over 1 m distances [Degen17, Pezzè16]. The potential impact includes infrastructure, climate research, geophysics and underground aquifer control, enhanced oil and mineral recovery, carbon storage and natural disaster pre-warning in the area of earthquakes and volcano activity.

**Quantum clocks.** Atomic clocks are the most established example of quantum technology, having been used since 1967 for international timekeeping. Optical clocks currently under investigation range from neutral atoms in optical lattices and singly charged ions and molecules to highly-charged ions and even nuclear transitions. Neutral atoms offer a high signal-to-noise ratio but are in general more susceptible to external fields and collisional shifts, requiring their environment to be well-controlled. In contrast, single ion setups can be very simple and technologically less demanding to achieve a similar level of accuracy as their neutral atom counterparts [Ludlow15] at the expense of longer averaging times. So far most of quantum clocks were limited by SQL, but first demonstrations of enhanced SNR through spin squeezing in microwave clocks have been reported [Kruse16, Leroux10].

**Quantum imaging.** Related to precision sensing using light is the idea of image acquisition. One analogy is that an image is a set of parameters that characterise the object, acquired in a massively parallel manner. This intrinsic feature of optical imaging enables exploitation of the different degrees of freedom of light: its spatial and temporal (or, equivalently, directional and frequency) structure, to enable optical resolution beyond the standard wavelength limit, with low light levels, or in the presence of strong background illumination. For instance, one proposed application is in quantum microlithography, where the quantum entanglement of the spatial degrees of freedom of light beams is able to affect matter at a scale smaller than the wavelength by patterning substrates by means of intensity correlations. Detecting details in images smaller than the wavelength has obvious applications in the fields of microscopy, pattern recognition and segmentation in images, and optical data storage. Correlations between quantum light beams enables new modes of imaging such as so-called "ghost imaging" in which an image of an object that is illuminated by one beam is acquired by a camera looking at a different beam, that did not impinge on the object.

## Spin-qubit-based sensing

Sensing using spin qubits is a relatively new and upcoming field in quantum sensing. While sensing magnetic fields comes most naturally for spin sensors [Balasubramanian08] and is of crucial importance for several fields for science including chemistry, biology, medicine and material science, spin-based sensing of a variety of different quantities, including temperature, electric field and pressure as well as force or optical near-fields has been demonstrated with diamond defect and defects in silicon carbide. All rely on the long living quantum coherence of spins to build robust, calibration free sensors. These devices operate by measuring the quantum phase accumulated by a qubit in the presence of the external perturbation. Coherent control of qubits including dynamical decoupling techniques is crucial for achieving best performance.

At present quantum spin sensors are targeting the following benchmarks: high sensitivity; spatial resolution; spectral and temporal resolution (when measuring AC fields). Note that high sensitivity and spectral resolution in quantum metrology requires long spin coherence times, which often is not compatible with room temperature operation for variety solid state qubits (crucial for applications in life sciences). Single spin qubits in diamond are outstanding in this respect, since the diamond lattice allows for millisecond coherence time of electronic spins even under ambient conditions.

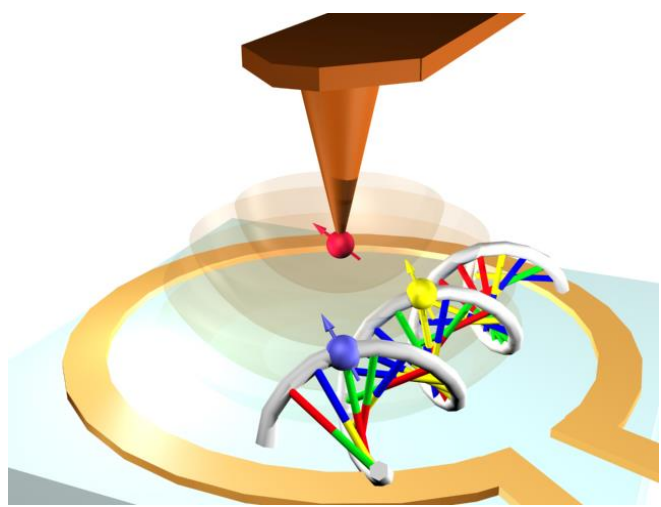

*Figure 5: Artistic depiction of a spin based quantum sensor for unravelling structure of single biomolecules*

## Optomechanical sensors

In the past decade, a technological and scientific paradigm shift has taken place around the optical and quantum control of nano- and micromechanical devices. NEMS (nano-electromechanical systems) and MEMS can now be measured and controlled at the quantum level by coupling them to optical cavities or superconducting microwave circuits. Recent demonstrations include squeezed mechanical states and QND measurements of mechanical motion, quantum coherent coupling in the optical and microwave domain, optomechanical ponderomotive squeezing and entanglement, a photon-phonon interface, and real time quantum feedback, among many others [Aspelmeyer14] . Current research in this field explores the physical limits of hybrid opto- and electro-mechanical devices for conversion, synthesis, processing, sensing and measurement of electromagnetic fields, from radio and microwave frequencies to the terahertz domain. The ability to modulate, interconvert, amplify or measure electromagnetic fields in this spectral region, is relevant to a number of existing application domains, specifically medicine (e.g., MRI imaging), security (e.g., Radar and THz monitoring) positioning, as well as timing and navigation (oscillators). At the same time, optomechanical systems provide an on-chip architecture to realize e.g., sensing, acceleration measurements, as well as low-noise amplification and novel non-reciprocal microwave components. While such devices can be used already in a classical context, where measurement of weak signals is relevant, extending the operation range into the quantum regime opens applications also in quantum science and technology, including quantum frequency translation from visible photons to the telecommunication band or realizing single-photon optical-to-microwave conversion, as well as sensors e.g., for charge, magnetic fields or mass. In addition, the ability to operate such optomechanical transducers in a regime where quantum noise plays a role also enables to create compact quantum noise calibrated thermometers.

## Advances in science and technology needed to meet future challenges

### Atomic and optical sensors

It remains a challenge for the field to demonstrate experimentally that it is possible to surpass the standard quantum or interferometric limits (SQL/SIL) in lossy sensors. In the case of photonic sensors, for example, it is known that the classes of quantum states that achieve this depend on the degree of loss, and that the Heisenberg scaling limit is never achieved when losses are present. Nonetheless, for all platforms, certain entangled states can give considerable improvements above the SQL. Squeezed states are certainly more robust for larger losses and have been used to improve the SNR in interferometric sensors, and for these states improving coupling of the probe to the sensor and reducing losses are key improvements. Atomic sensors typically suffer from lower losses than photonic sensors, but are more subject to dephasing noise. For cold atomic ensembles, the ability to prepare the initial probe states limits the repetition rate of the sensor, whereas for hot ensembles atomic motion is the limiting factor. In both cases, chip-scale integration will be important for space, mobile and personalized sensors. Further, combining photonic and atomic platforms may yield new capabilities [Wolfgramm13]. The instability of all optical clocks is currently limited by the residual noise of the clock laser. The challenge is to further improve existing techniques for laser frequency stabilisation based on e.g. cavities, spectral hole burning, or even lasing on a clock transition. New clock technology needs to be combined with reductions in size, weight, power consumption and cost.

Theoretical study of quantum sensing remains a critical element in order to examine the fundamental limits of metrology. Theory will help to inform the experimentalist how much more effort needs to be expended to attain the known bounds. In particular, new measurement protocols as well as post-processing of the measurement outcomes can be further optimized. For instance, feedback-based protocols, dynamical decoupling, and optimal control may all add new capabilities to quantum sensing protocols. Powerful methods from signal processing, which have already yielded fruit in the design and assessment of sensor performance, could be applied to minimize the measurement effort to extract the desired signal.

### Spin-qubit-based sensing

Although first proofs of principle demonstration show high potential of diamond sensing devices for magnetic field sensing, key challenges than need to be addressed in order to bring this technique to application is integration in user-friendly prototype. Depending on the application, this comprises optical integration and combination with control electronics. For medical and bio-analytical applications, integration into existing analytical devices like fluorescence microscopes is needed.

Quantum control tools open new technique that will improve sensitivities and open new application areas. So far, quantum entanglement between spins remained widely unexplored. For example, concentration of NV centres for ensemble NV magnetometry was adjusted to be low enough to avoid dipole–dipole coupling between spins. On the other hand, such coupling provides an opportunity to generate squeezing in dense spin systems and reach sensitivities approaching the Heisenberg limit.

Applications of NV magnetometers in life sciences and medicine depend on the ability to insert nanodiamonds doped with colour centres into cells. Sensing can be combined with other functionalities of nanodiamonds (for example their use as drug delivery devices or markers for ultra-sensitive MRI enabled by hyperpolarisation of nuclear spins). A remaining challenge is the size reduction of nanodiamonds as well as their versatile surface functionalisation allowing selective protein targeting.

## Optomechanical sensors

Materials and fabrication challenges have a strong bearing on current optomechanical devices. A significant medium-term challenge is to fabricate hybrid nano-optomechanical systems in combination with standard CMOS processing, thereby making them compatible with current manufacturing methods. Reducing optical losses will allow on-chip architectures to exploit full quantum control, e.g., via coherent feedback, perform full quantum state tomography, etc. In turn, this will allow preparation of quantum states that are known to improve sensing and transduction sensitivity. Lower-absorption materials are also crucial in reducing the thermal load on devices. In combination with a wide variety of different methods, including pulsed protocols, using squeezed light, etc., this would help to extend the quantum regime to lower frequencies and larger masses, which enables broader sensing capabilities. Alternative routes to drastically reducing mechanical dissipation include the use of phononic band-gap architectures and substrate-free levitated topologies, which will eventually allow quantum operation at room temperature.

## Conclusion

The potential impact of quantum sensors is broad and considerable. A variety of different platforms enables quantum-enhanced measurement of time, space, rotation, as well as gravitational, electrical and magnetic fields. All these technologies find important applications in fields as physics, chemistry, biology, medicine or data storage and processing.

We thank the many members of the community who have contributed to the content of this article.

# Quantum Control

Frank K. Wilhelm[13], Steffen J. Glaser[14]

## Introduction

It is control that turns scientific knowledge into technology. The general goal of quantum control is to actively manipulate dynamical processes of quantum systems, typically by means of external electromagnetic fields or forces. The objective of quantum optimal control is to devise and implement shapes of pulses of external fields or sequences of such pulses, that reach a given task in a quantum system in the best possible way. Quantum control builds on a variety of theoretical and technological advances from the fields of mathematical control theory and numerical mathematics all the way to devising better electronic devices such as arbitrary-waveform generators.

The challenge to manipulate nature at the quantum level offers a huge potential for current and future applications both in traditional applications and in modern quantum technologies. It is part of the effort to engineer quantum technologies from the bottom up, and many striking examples of surprising and non-intuitive - but extremely efficient and robust - quantum control techniques have been discovered in recent years. While the precise way to manipulate the behaviour of these systems may differ from ultrafast laser control to radio waves, the control, identification and system design problems encountered share commonalities, while at the same time being distinct from classical control problems.

The European quantum control community has come together in the FP 7 coordination action QUAINT that persists to be connected through the website www.quantumcontrol.eu. The community has written its own roadmap [Glaser15] which is very detailed and covers both first- and second generation quantum technologies.

## Current status

Quantum control theory is addressing two fundamental questions, that of *controllability*, i.e., what control targets are accessible and that of *control design*, i.e. how can a target be reached. Approaches for control design can be open-loop or closed-loop. In the latter case, the specific nature of quantum measurements needs to be taken into account. Open loop techniques include approaches based on the Pontryagin maximum principle, i.e., quantum optimal control, with solutions obtained analytically or numerically. Optimal control theory does not make any restrictive assumptions on the quantum system and also experimental constraints and robustness requirements can be fully taken into account (the latter is called simultaneous controllability) and is hence broadly applicable. Closed loop techniques involve the use of feedback to stabilize a given system state or to obtain a desired quantum input-out gain. As in classical engineering, the mathematical problem is controller design. In the quantum situation this can be measurement-based or fully coherent [Gough09].

Currently, the theory of controllability is well and rigorously understood for closed systems with finite-dimensional state space and there is solid understanding of the Markovian open case as well as a few results outside those paradigms. Analytical solutions are available for simple, low-dimensional as well as pseudoadiabatic systems. Although numerical approaches such as gradient ascent, Quasi-Newton, Newton and Krotov methods have reached a reasonable maturity and lead to robust and tailored software packages

[13] Theoretical Physics, Saarland University, D-66123 Saarbrücken, Germany
[14] Department of Chemistry, Technical University of Munich, Lichtenbergstrasse 4, D-85747 Garching, Germany

[Khaneja05, Reich12, Machnes11], many opportunities exist to significantly improve their performance. They are complemented by gradient-free approaches including the chopped random basis (CRAB) method [Doria11, Egger14]. Important challenges include increasing the speed of algorithms, broadening the base of controllability research and to integrate these techniques with a broader base of platforms.

Quantum optimal control is standing on the shoulders of its early applications in standard nuclear magnetic resonance (NMR) spectroscopy and atomic physics. They have pioneered standardization and software packages and currently pursue robust ensemble control as a central goal, which is also important for maturing second-generation quantum technologies as many of the challenges in quantum sensing and computing are closely related [Dolde14].

Successful implementation of quantum technologies needs to be carried out with sufficient accuracy, despite imperfections and potentially detrimental effects of the surroundings. Quantum optimal control toolboxes allow to identify the performance limits for a given device implementation and show how to reach those limits of operation. In order to obtain these results, the quantum optimal control methodology has been adapted to the requirements of Quantum Technologies, specifically including open system effects and optimizing for quantities like entanglement capabilities directly. They were adapted to nonlinear dynamics as found in BECs.

Quantum optimal control is related to information theory. It provides a practical means to explore decoherence-free subspaces or other noise-avoiding strategies as well as cooling schemes needed, e.g. to motionally cool levitating superconducting spheres. It is also related to quantum engineering by providing a solid mathematical framework for some engineering tasks. These include control of open systems and coherence control such as enhancing the lifetime of quantum memories by dissipative state engineering. More globally, both together aim at the convergence of optimal control and experimentation including calibration uncertainties and other constraints.

## Advances in science and technology needed to meet future challenges

A key family of mid-term challenges to optimal control is to improve and reach convergence between theory and experiment in more platforms than previously. With this, control methods will be crucial to operate these devices reliably and accurately. This involves the device preparation or reset, the execution of the desired time evolution, and the readout of the result.

In the long run, when scaling quantum technology, control needs to scale with it. Meeting this challenge is necessary for proper functioning in a world that is only partially quantum. Next to finding these controls, benchmarking their success will be of nearly equal importance.

**Applications in quantum communication:** Quantum communication connects to quantum optimal control mostly at the light-matter interface. Currently, many proposals for transport as well as photon storage were made. Going forward, quantum control will develop into schemes to stabilize networks with feedback and optimize interconversion between stationary and flying qubits.

**Applications in quantum computation:** The ongoing theme here is the optimal design of powerful gates and state preparation schemes. Single-qubit gates were made robust against frequency crowding and slow fluctuations, even in complex Hilbert spaces and control schemes were constructed that make active use of environmental degrees of freedom. This needs to be driven towards robustness even in multi-qubit architectures and to the case of large inhomogeneity as common in semiconductor spin qubits. Going to optimal two-qubit gates, optimal control helps finding faster strategies solving the platform-specific challenges of high fidelity, error correction, long-distance entanglement, and robustness. Optimal control also needs to improve performance of qubit measurement and reset. Speeding up gates and combinations of gates and transport will remain a challenge. With promising starts in closed-loop fine tuning in superconducting qubits, the automation of control design and its integration with error correction as processors are scaled needs to be further developed. In the long run, optimal control is a crucial ingredient for quantum compilers and a scalable language for the assembly of elementary or complex gates in multi-qubit systems. Next to the gate-based model of quantum computing, quantum optimal control proposals for preparing cluster states have been made and can be extended.

**Applications in quantum simulation:** Quantum simulation is proving to be a flexible and inspiring field for applications of quantum optimal control, e.g. in the platform for quantum simulation in optical lattices [Rosi13]. There, it has contributed to improved loading of atoms and found serendipitous solutions for local control. This should be broadened into the optimal and robust creation of more complex entangled states both for this and for other simulation platforms. They can be taken out of equilibrium to help study the emergence of thermodynamic laws, e.g. for spin systems, proposals for preparation of many-body entangled non-classical states were made.

For quantum simulation as special purpose quantum computing, optimal control helps explore fidelity limits in the presence of noise, both Markovian and non-Markovian as it occurs, e.g., in collision models. It will be used to keep control and operation fidelity high during the aggressive scaling anticipated in simulation and in the long-run be pivotal in verifying and validating simulations that are performed without or with limited error correction.

**Applications in quantum sensing:** Starting from its foundation in NMR, see above, quantum optimal control is naturally applied to quantum sensing. For example, the concurrent optimization of pulses with the ability to cancel each other's imperfections was demonstrated to yield ultra-broadband Ramsey experiments (see Fig. 6).

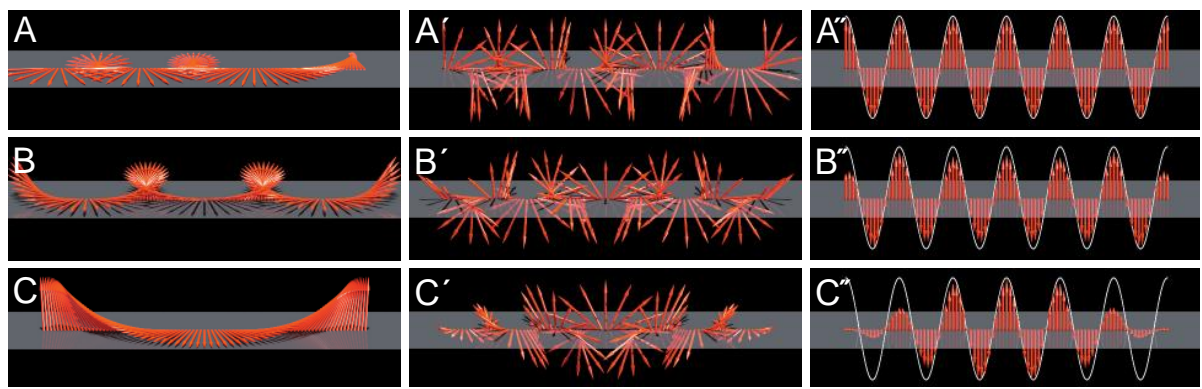

*Figure 6: Offset-dependence of the Bloch vector during the course of a Ramsey experiment using three different pulse sequences with the same maximum amplitudes: (A-A'') concurrently optimized broadband excitation and flip-back pulses that cancel each other's imperfections in a cooperative fashion, (B-B'') individually optimized broadband pulses of the same duration, and (C-C'') standard rectangular pulses. The offset-dependent orientation of the Bloch vector is shown after the excitation pulse (left panels) and after a delay followed by a flip-back pulse (center panels). The right panels show the corresponding z component of the final Bloch vector and the white curves represent the desired ideal Ramsey fringe pattern (adapted from Braun and Glaser, New J. Phys. 16, 115002, 2014).*

Protocols for sensing using spins of NV centres in diamond were already developed and are expected to be further improved to protect from noise while enhancing the signal both by improving decoupling and preparing squeezed states. Non-classical states are a key ingredient to sensing and were also proposed for BECs [vanFrank14] and photons in a cavity. A further application challenge in optimal control for sensing is to use feedback and adaptive settings for extracting phases and other parameters in the best way possible.

## Conclusion

The long-term goal of quantum optimal control for quantum technologies is to gain a thorough understanding of optimal solutions and to develop a software layer enhancing the performance of quantum hardware for tasks in computing, simulation, communication, metrology and sensing beyond what is achievable by classical means, enabling the achievement of quantum supremacy.

We thank the many members of the community who have contributed to the content of this article.

# Quantum Software and Theory

Antonio Acín[15], Harry Buhrman[16]

## Introduction

Computers connected through networks, as we know them today, have changed modern society fundamentally, but their development is far from over. In fact, we are just starting to harness the laws of quantum physics to process information in unprecedented ways. Since the development of the first quantum algorithms and protocols there has also been steady and impressive progress on the hardware side, delivering quantum systems with a small number ($< 20$) of qubits and quantum networks ranging over several hundreds of kilometres. With this progress, the need for quantum software and theory to exploit these novel quantum technologies, understanding their power but also their limitations, becomes more and more urgent. In the following sections, we highlight the current status and future challenges of this theoretical effort structured along three main research directions: quantum software for computing, quantum software for networks and theory.

## Current Status

**Quantum software for computing:** Quantum algorithms are fundamentally different from their classical counterparts because qubits can be in a superposition of 0 and 1. This means that with n qubits one can potentially perform exponentially many ($2^n$), computations in parallel. However, it is difficult to extract the answer from such a superposition as observing the system collapses it. This is where quantum software is needed. Shor showed in 1994 [Shor94] that numbers can be factored more efficiently by quantum computers (QCs), an immensely important discovery given that the security of many modern cryptographic protocols (such as RSA) are based on the assumption that factoring large integers is a computationally hard task. Other algorithms were developed for a wide range of problems such as for example searching, sorting and many other applications [Montanaro16]. One of the first practical applications of QCs may be quantum simulation [QSim12], as even modest devices have the potential to perform simulations that would be infeasible with classical computers. There exist physical systems in which the interactions necessary for simulation can be engineered without the need for a full QC. With 100 to 150 logical qubits, molecular energies can be computed to great precision and accuracy, far exceeding the limitations of classical computers. Carrying out coherent quantum operations despite noise is a key challenge. Active strategies (error-correcting codes [Lidar13]) as well as passive ones (error-avoiding codes) have been introduced. Recent developments have reduced the noise threshold estimate for quantum error correction by several orders of magnitude. Topological quantum computation encodes quantum states and gates in global, delocalised properties of the hardware medium, which are more immune to all forms of noise that do not impact the entire medium at once and coherently. Protocols for the certification of correct quantum computation become essential in all these setups. Methods to test arbitrary computations with little overhead have been proposed, as well as other approaches to test quantum computers based on interactive-proof systems. On the other hand, new algorithms for the efficient simulation of quantum models have been developed, for instance based on tensor-network techniques. Finally, different architectures for quantum computation have been proposed, such as the gate or circuit model, adiabatic quantum computing, and quantum cellular automata, among others.

[15] ICFO-Institut de Ciencies Fotoniques, The Barcelona Institute of Science and Technology, 08860 Castelldefels (Barcelona), and ICREA, Passeig de Ll. Campanys, 23, 08010 Barcelona, Spain

[16] QuSoft, CWI, and University of Amsterdam, Sciencepark 123 1098 XG, Amsterdam, The Netherlands

**Quantum software for networks:** Just as quantum algorithms can lead to an exponential speed-up for computational problems, quantum communication can lead to exponential savings in the number of (qu)bits that must be transmitted to solve distributed problems [Buhrman10]. Some of these protocols have already been implemented, such as the quantum-fingerprinting scheme and the vector in a subspace problem. Cryptographic protocols also take place on networks and quantum resources allow, for certain problems, security guarantees that are impossible to achieve classically. Quantum key distribution (QKD) [Bennett84], for instance, allows two mutually trusting parties to generate a shared secret key. QKD systems are already commercially available. Cryptographic tasks where the sender and receiver do not trust each other require additional assumptions, limiting the adversary's computational or physical power. In the first case, there are quantum proposals for quantum cloud computation (blind computation), quantum money, and position-based cryptography. Limiting the adversary's physical power, i.e. amounts of quantum memory or entanglement or guaranteed space-like separation between participants, leads to a broad range of protocols which are easy to implement on existing hardware. Another line of research is quantum-safe or post-quantum cryptography where protocols are proven to be secure based on the hardness of certain problems, such as lattice problems. To make optimal use of quantum networks it is required to understand how to distribute quantum resources over them. Recently, there have been a few breakthroughs with respect to the non-additivity of quantum and classical information capacity and the key problem of identifying information capacities has been solved for a significant subset of channels. Protocols for entanglement distribution are necessary for long-distance quantum communication and the vision of a quantum internet [Kimble08]. As for computation, certification methods have also been introduced in the context of networks, for instance to certify the presence of entangled states or the security of quantum channels.

**Quantum information theory:** as its classical counterpart, quantum information theory aims at identifying the laws and the ultimate limits governing any information process based on quantum effects. Theoretical frameworks known as resource theories have been developed to understand quantum resources, such as entanglement [Horodecki09], non-locality [Brunner14], quantum randomness or secret bits. Efficient strategies to estimate relevant quantum properties have also been designed. From a fundamental perspective, these concepts have been applied to understand what makes quantum physics special, devise novel no-go theorems for classical simulation of quantum physics or the quantum-vs-classical transition, also necessary to understand decoherence. Finally, quantum information concepts have successfully been applied to other domains in science, such as many-body physics [Amico08], quantum chemistry and biology, quantum thermodynamics, quantum gravity, high-energy physics and even to solve open problems in classical information and computation theory.

## Advances in science and technology needed to meet future challenges

**Quantum software for computing:** A constant challenge in this field is to find new quantum algorithms that outperform the best classical algorithms. However, quantum algorithms cannot yield an advantage for every problem; in fact, they usually do not, and understanding also these limitations will be critical, for instance in developing quantum-resistant classical and quantum cryptography, or to derive no-go theorems for quantum computation. Most of the existing algorithms do not make reference to any specific implementation and often cannot be implemented on the 50–100 qubit platforms available in the medium term. In the coming years, new algorithms and applications will be developed for these small platforms with a limited number of qubits where classical simulation is impossible, aiming at demonstrating "quantum supremacy". In this direction, it is important to understand how these medium-size quantum processors can be used to simulate systems of physical relevance, for example in quantum chemistry, material science or high-energy physics. Assessing the impact of errors on computation quality remains a challenge and will require more efforts. In standard computation, new schemes for error correction and fault-tolerant computation, including ideas from topological quantum computation, need to be designed so that the level of noise that can be tolerated under realistic error models in near-future quantum systems is increased. In simulation, the impact of errors needs to be understood: while a single error in a QC without error correction is fatal, a small error in, say, a measurement of conductivity is less critical. Certifying a given quantum computation when a classical simulation is impossible represents another challenge and here improved algorithms for the classical simulation of quantum processes will be essential. Finally, first steps in extending machine learning and artificial intelligence applications to the quantum realm have taken place and it is expected that more algorithms will be designed in the next years.

**Quantum software for networks:** Finding new protocols for distributed computation also remains a challenge. For that, we need to understand the power that the entanglement-assisted communication model offers. Here, it will be again important to understand how to adapt existing or design new protocols for the near-future implementations. Concerning QKD, the development of device-independent techniques is essential to design implementations robust against existing hacking attacks. A major theoretical, and also experimental, challenge is to make these proposals practical. Recently, loophole free Bell tests have been achieved, but further work is required to speed up the rate at which we could hope to generate a key in QKD. It is also important to extend cryptographic applications beyond QKD. Improvements should be expected in the design for protocols involving non-trusted parties, which usually require computational or physical assumptions. In general, limiting the adversary's physical power, i.e. amounts of quantum memory or entanglement or computational power, will lead to a broad range of protocols which are easy to implement on existing or near-future hardware. Remaining challenges include more complicated tasks such as secure identification. We also expect more efficient protocols for post-quantum cryptography. Finally, more work is needed to optimise the quantum resources for communication over quantum networks. Further investigation is needed to identify similarities and differences between classical and quantum network theory, and to consider practical constraints like channel uncertainty, finite block size, and limited entanglement.

**Quantum information theory:** To understand the full power of quantum effects, instrumental theories for quantum information resources, such as number of qubits, entanglement, various aspects of secrecy, study of randomness or channel capacities will need to be developed. Assessing the successful implementation of quantum protocols will require the design of efficient and scalable methods for the estimation, detection and certification of quantum properties. We also expect quantum information concepts and techniques to have impact on other research fields. A quantitative theory of entanglement could provide new insights into the exact structure of correlations of many-body systems, possibly leading to new algorithms for their simulation. This may lead to the identification of novel phases of matter from a quantum information perspective and for quantum information purposes. The role of quantum coherences in biological and thermodynamic processes also requires further investigation.

We expect that some important headway will be made by the challenges and milestones above within the next five years. In particular implementations on small quantum systems as they become available. Also new schemes for error correction and fault-tolerance amenable to such small systems. With additional manpower and new insights, it is also expected that new quantum algorithms will be developed within the next 5 years.

## Conclusion

Software, protocols, and quantum information theory are essential for an optimal development of quantum technologies. Until now, most of the effort has focused on identifying the ultimate limits for quantum information processing. In the next 5-10 years, a parallel effort will be devoted to understand what can be done with the first generations of small quantum processors, identifying for instance quantum computation protocols whose classical simulation is infeasible or realisation of protocols with unprecedented levels of security. In the long term, these two efforts are expected to converge, providing the tools to attain the ultimate limits for quantum information processing with the, by then, existing technologies.

We thank the many members of the community who have contributed to the content of this article, in particular I. Cirac, M. Troyer, S. Wehner, R. Werner, A. Winter, and M. Wolf.